Commentary

# The Role of Intelligent Transportation Systems and Artificial Intelligence in Energy Efficiency and Emission Reduction

Omar Rinchi, Ahmad Alsharoa[1],
Ibrahem Shatnawi, and Anvita Arora[2]

[1] Omar Rinchi and Ahmad Alsharoa are affiliated with the Department of Electrical and Computer Engineering, Missouri University of Science and Technology, Rolla, MO 65409, USA.

[2] Ibrahem Shatnawi and Anvita Arora are affiliated with the King Abdullah Petroleum Studies and Research Center (KAPSARC), Riyadh, Saudi Arabia.

## About KAPSARC

KAPSARC is an advisory think tank within global energy economics and sustainability providing advisory services to entities and authorities in the Saudi energy sector to advance Saudi Arabia's energy sector and inform global policies through evidence-based advice and applied research.

This publication is also available in Arabic.

## Legal Notice

# Abstract

Despite the technological advancements in the transportation sector, the industry continues to grapple with increasing energy consumption and vehicular emissions, which intensify environmental degradation and climate change. The inefficient management of traffic flow, the underutilization of transport network interconnectivity, and the limited implementation of artificial intelligence (AI)-driven predictive models pose significant challenges to achieving energy efficiency and emission reduction. Thus, there is a timely and critical need for an integrated, sophisticated approach that leverages intelligent transportation systems (ITSs) and AI for energy conservation and emission reduction. In this paper, we explore the role of ITSs and AI in future enhanced energy and emission reduction (EER). More specifically, we discuss the impact of sensors at different levels of ITS on improving EER. We also investigate the potential networking connections in ITSs and provide an illustration of how they improve EER. Finally, we discuss potential AI services for improved EER in the future. The findings discussed in this paper will contribute to the ongoing discussion about the vital role of ITSs and AI applications in addressing the challenges associated with achieving energy savings and emission reductions in the transportation sector. Additionally, it will provide insights for policymakers and industry professionals to enable them to develop policies and implementation plans for the integration of ITSs and AI technologies in the transportation sector.

# Introduction

The transportation sector in Saudi Arabia plays a significant role in contributing to carbon emissions and other environmental issues. Traditional private cars are the primary mode of transportation in the country, accounting for a significant portion of greenhouse gas emissions (United Nations 2014). The use of such vehicles has led to traffic congestion and air pollution, which in turn can lead to an increase in energy consumption. To address these challenges, there has been a growing need to promote the use of advancements in technology and data science, demonstrated by the integration of intelligent transportation systems (ITSs) and artificial intelligence (AI), which can mitigate the problems posed by energy demand and greenhouse gas emissions.

ITSs refer to a vast network of cutting-edge technologies and applications created to improve the performance of transportation systems in terms of sustainability, efficiency, and safety (Veres and Moussa 2020). Using information and communication technology, ITSs contain several components, such as people, vehicles, devices, and infrastructure. More specifically, ITSs encompass numerous interconnected architectural components, as shown in Fig. 1. Real-time data are collected by sensors and data sources that are installed in vehicles, devices, and infrastructure. Data sharing between components is made possible by communication networks, including backhaul, vehicle-to-vehicle (V2V), and vehicle-to-infrastructure (V2I) networks. To make efficient decisions and manage the transportation network, control centers need to process and evaluate the received data. Intelligent apps and algorithms can process the data to optimize traffic flow, increase safety, and offer environmentally friendly services. Travelers can access real-time information and services through user interfaces and applications, enabling them to select the optimal route for their journey and thus contributing to reducing energy consumption and emissions.

**Figure 1.** ITS Infrastructure and Communication Architecture.

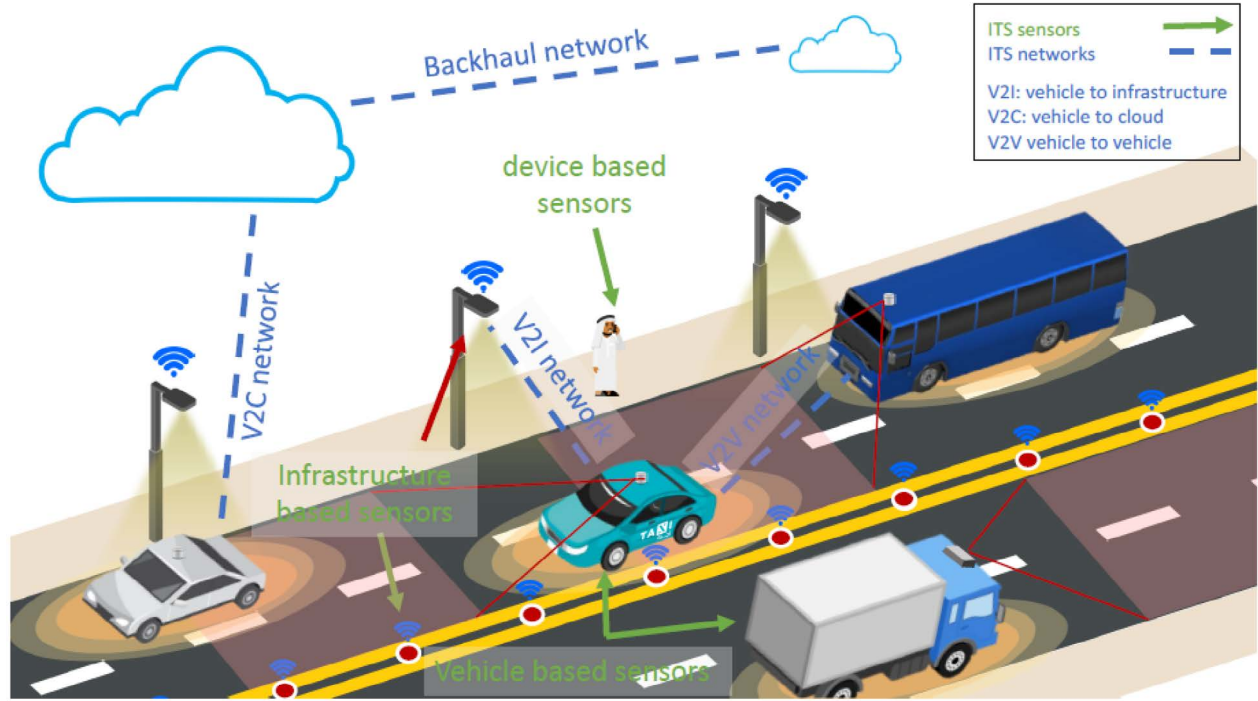

*Source: Authors.*

# ITS Sensor-Based Environmentally Friendly Services

The range and depth of services that an ITS can offer are intrinsically linked to its sensing capabilities, forming a dynamic ecosystem where enhanced sensing facilitates increased service delivery. We categorize ITSs from the sensor level into three categories: (i) vehicle-based sensors, (ii) infrastructure-based sensors, and (iii) device-based sensors. With proper optimization of ITS sensing capabilities, it is possible to achieve nontraditional enhanced energy and emission reduction (EER) services such as eco-driving assistance, traffic flow optimization, and intelligent freight management.

**Vehicle-based sensors:** Vehicle-based sensors encompass a diverse range of sensors integrated within vehicles to collect and monitor various types of data (Guerrero-Ibáñez, Zeadally, and Contreras-Castillo 2018). These sensors can be categorized as follows based on their overall goals within an ITS.

- Surrounding/environmental sensors, such as light detection and ranging sensors (LiDARs), cameras, temperature sensors, humidity sensors, and air quality sensors, capture information about a vehicle's surroundings, road conditions, and environmental parameters. These sensors enable applications such as object detection, lane departure warning, weather monitoring, and pollution detection.

- Vehicle-specific sensors, including speed sensors, brake sensors, engine sensors, fuel consumption sensors, emission sensors, energy meters, cameras, and LiDARs, monitor parameters related to a vehicle's performance, energy efficiency, emission levels, and energy consumption. These sensors facilitate vehicle diagnostics, fuel economy analysis, compliance with environmental regulations, and energy management.

- Vehicle-based sensors provide data on vehicle performance, fuel consumption, and emissions. These data can be used to optimize driving behavior through eco-driving assistance systems, minimizing fuel consumption and reducing emissions. Additionally, vehicle sensors enable the monitoring and management of electric vehicles, promoting the adoption of zero-emission vehicles and reducing the reliance on fossil fuels.

**Infrastructure-based sensors:** Infrastructure-based sensors encompass a diverse range of sensors that are strategically deployed within transportation infrastructure to fulfill specific functional roles in an ITS (Soga and Schooling 2016). Infrastructure-based sensors and their uses are listed below.

- Road condition monitoring sensors, such as pavement quality analyzers and surface temperature sensors, are installed directly on the road surface to continuously assess pavement conditions, detect cracks and potholes, and monitor temperature variations.

- Vehicle presence and behavior sensors, including license plate recognition cameras, LiDARs, and

- weigh-in-motion systems, are strategically positioned on traffic lights, overhead structures, or roadside gantries to capture crucial data related to vehicles on the road, such as identification, classification, speed, and weight measurements. These sensors facilitate efficient traffic management and enforcement operations, and they support autonomous vehicle (AV) technologies.

- Environmental monitoring sensors, such as weather stations, air quality sensors, visibility sensors, and noise sensors, are strategically located near roadways or in surrounding areas to monitor weather conditions, air pollution levels, visibility, and noise levels, aiding informed decision making for climate adaptation, pollution control strategies, and urban planning.

- Traffic monitoring sensors, such as inductive loop detectors embedded in the road surface, microwave radar sensors, Bluetooth detectors, and cameras, enable real-time traffic flow analysis, vehicle counting, speed measurements, and incident detection (Klein, Mills, and Gibson 2006). These sensors are strategically placed at specific points within the road network to provide vital information for congestion mitigation, signal optimization, and proactive traffic management.

At the infrastructure level, sensors such as traffic sensors and environmental sensors contribute to traffic flow optimization and congestion reduction. By efficiently managing traffic signals based on real-time data, these sensors help minimize the number of stops and delays, leading to a smoother traffic flow and reduced fuel consumption (Bernas et al. 2018). Environmental sensors assist in monitoring air quality, weather conditions, and visibility, providing insights for pollution control measures and climate adaptation strategies.

**Device-based sensors:** Device-based sensors refer to sensors integrated within personal devices, such as smartphones, wearables, and connected devices. These sensors capture various types of data, including location from global positioning system (GPS) sensors, movement from accelerometers and gyroscopes, proximity and light information from sensors, and environmental data such as temperature and humidity. Image sensors, such as cameras, allow for visual data collection.

The data obtained from these sensors have diverse applications in ITSs, including real-time traffic monitoring, travel analysis, personalized navigation, crowd sensing for road conditions, and incident reporting. Device-based sensors can also gather user-specific data such as biometrics for driver monitoring and health-related applications. Device-based sensors offer opportunities for personalized services and real-time data collection. By utilizing location data, travel behavior patterns can be analyzed to optimize transportation services, encourage shared mobility options, and reduce the number of individual vehicle trips. Moreover, these sensors can facilitate user engagement and behavioral change through feedback and recommendations, encouraging energy-efficient transportation choices (Nafrees et al. 2021; Rinchi, Assaid, and Khasawneh 2021).

Optimizing sensing resources requires an integration of sensing capabilities from different levels. For instance, in Fig. 2, sensors on both the vehicle and infrastructure are appropriately utilized to enhance EER. More specifically, the vision sensors on the traffic light (e.g., cameras, LiDARs, or depth sensors) send real-time information about traffic flow, traffic light timing, and road obstructions, while GPSs, inertial measurement units (IMUs), and vision sensors on the vehicle send more insights about the position, speed, and surroundings of the vehicle, respectively (Ma et al. 2009). An algorithm located on an edge device, the cloud, or a central processor that could take the form of a deterministic algorithm or AI processes the received data and sends back navigation recommendations for each vehicle, while traffic lights receive timing and management commands. Such a scenario can directly pave the way for advanced traffic management strategies such as cooperative adaptive cruise control (CACC) and smart intersection management, which can result in a significant improvement in EER. Furthermore, intelligently adjusting traffic light timings and vehicle navigation recommendations can reduce the idling time at intersections. Less time spent idling leads to decreased fuel consumption and, consequently, lower emissions. An appropriate adjustment of vehicle speed and following distance based on current and anticipated traffic conditions can prevent unnecessary acceleration and braking, resulting in a smoother traffic flow and minimizing fuel consumption. Lower fuel consumption directly translates into lower emissions. Furthermore, these systems encourage smoother driving behaviors, reducing the need for rapid accelerations and sudden braking, which are inefficient in terms of fuel consumption and increase the wear and tear on vehicles. By promoting more uniform speeds and smoother driving, these systems further enhance energy efficiency and reduce emissions.

**Figure 2.** Hybrid sensing collaboration.

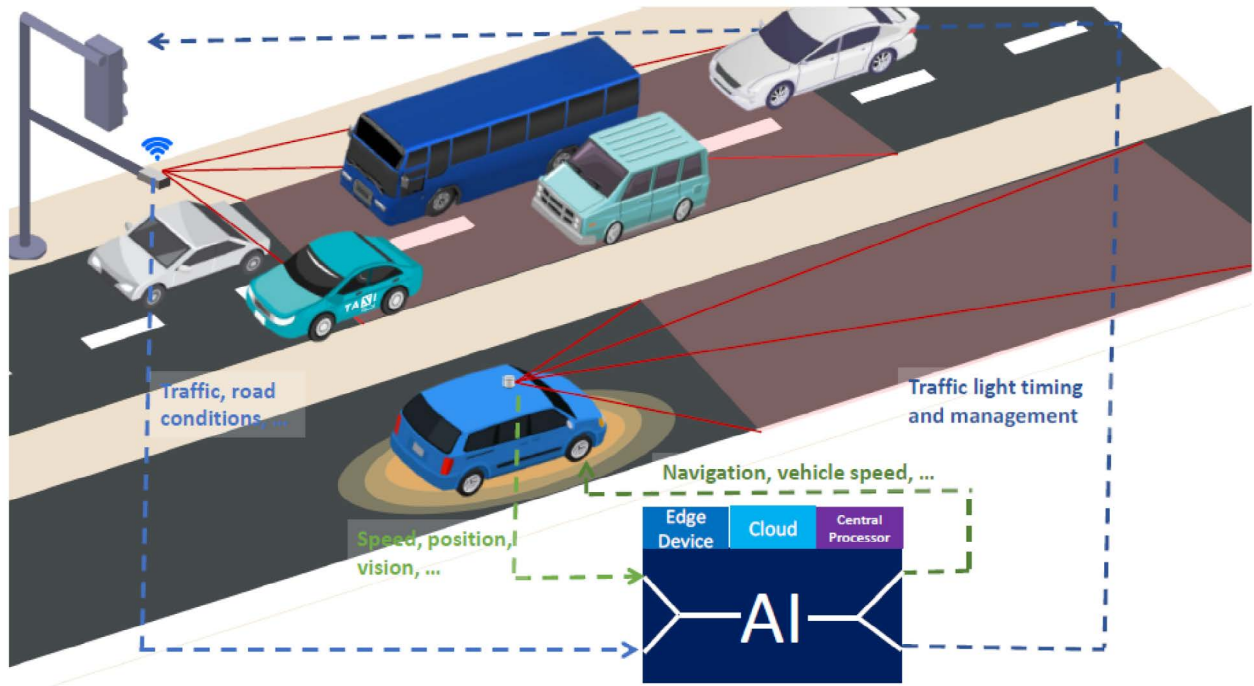

*Source: Author.*

The potential applications of integrated sensors in future transportation systems go beyond the enhanced EER problem, as they can yield more generic advanced ITSs. For instance, vehicle-based sensors enable real-time updates through consistent data collection and analysis, resulting in mitigated traffic congestion and shorter travel times. Simultaneously, they greatly enhance public transport efficiency by tracking vehicle locations, occupancy, and estimated arrival times.

In contrast, infrastructure-based sensors revolutionize smart parking and public transport management. They allow for effective space management, reduce the time spent searching for parking, and lower fuel consumption and emissions. Moreover, these sensors facilitate more efficient scheduling, route planning, and fleet management. Device-based sensors play a transformative role in user-facing applications such as smart ticketing and mobility as a service (MaaS). They allow for smooth, contactless transactions, reducing wait times and creating a more frictionless interaction with the transportation system. In the MaaS context, these sensors enable a seamless interface among various modes of transport, contributing to a more integrated and efficient urban mobility ecosystem. Finally, within ridesharing applications, sensors of all types ensure the accurate tracking of vehicles and passengers, providing a safe and reliable service.

New generations of vision sensors, such as LiDARs, are capable of converting the surrounding environment into a three-dimensional (3D) representation of point clouds (Rinchi et al. 2023). Unlike traditional red-green-blue (RGB) pixel-based video cameras, LiDARs do not collect personal identifiable information (PII), which protects the privacy of humans in their surroundings. Furthermore, LiDARs do not require any illumination to work, which allows them to work day and night. Moreover, due to their unique independence of texture in their surroundings, LiDARs are less subject to model discrimination for people with different skin colors, are less subject to optical illuminations, and can lead to easier background subtraction compared to other vision techniques. For these reasons, new directions are needed to integrate LiDARs into future ITS infrastructure, such as typical on-vehicle LiDARs, elevated LiDARs (ELiDs) (Lucic et al. 2020), and LiDAR-integrated unmanned aerial vehicles (UAVs) (ULiDs) (Osterwisch et al. 2023).

# ITS Interconnected Networks for EER

Modern transportation systems strive for higher energy efficiency and lower emissions, and the necessity of sharing data and commands in real time has increasingly grown. Therefore, an intricate network system composed of interconnected V2V, V2I, vehicle-to-cloud (V2C), and backhaul networks plays a necessary role (Khan et al. n.d.). Each network not only enables the real-time exchange of crucial information and commands but also serves unique functions in the quest to enhance transportation efficiency and reduce environmental impacts.

**V2V networks**: V2V communication networks utilize wireless technology for vehicles to share important operational data directly with each other. These systems often use dedicated short-range communication (DSRC) or cellular-V2X (C-V2X) technologies. DSRC, which is a form of Wi-Fi, operates in the 5.9 GHz band specifically set aside for intelligent transportation systems (Kamal, Srivastava, and Tariq 2021). In contrast, C-V2X uses cellular networks (4G or 5G) and operates in the bands assigned to the cellular provider or in the 5.9 GHz band. By maintaining a steady flow of data on each vehicle's speed, position, and direction, V2V communication enables vehicles to anticipate and respond to each other's actions, reducing the likelihood of accidents. In turn, this reduction in the likelihood of accidents ensures a smoother traffic flow, leading to more efficient fuel usage and lower emissions.

**V2I networks:** V2I communication networks enable vehicles to communicate with transport infrastructure, such as traffic signals, road signs, and traffic management systems. Similar to V2V, V2I communications also often use DSRC or C-V2X technologies (Gupta et al. 2022). V2I networks provide valuable information on traffic congestion, road conditions, and traffic light status, among others, facilitating real-time decision making to optimize routes and manage speeds. By minimizing stop-and-go traffic and unnecessary idling, V2I networks significantly contribute to enhancing fuel efficiency and reducing emissions.

**V2C networks:** V2C communication networks connect vehicles with various cloud-based services through wireless cellular networks, often using 4G long-term evolution (LTE) or 5G technology (Yu et al. 2022). This connection provides access to an extensive range of data, including real-time traffic updates, weather conditions, and predictive maintenance alerts. V2C plays a crucial role in efficient route planning and in optimizing vehicle performance, both of which contribute to lowering fuel consumption and reducing $CO_2$ emissions. Moreover, with advancements in machine learning (ML) and AI, cloud-based data can be leveraged to develop sophisticated traffic management systems and optimize vehicle performance.

**Backhaul networks**: Backhaul networks form the backbone of any ITS, acting as the essential conduit connecting peripheral components, such as sensors and roadside units (RSUs), to the core system where data processing and decision making take place. The heart of this system comprises several key elements. Access points (Aps) serve as the initial data collection nodes; for instance, in a V2I scenario, an AP could be an RSU amassing data from passing vehicles. These Aps connect to the central system through backhaul links, which can be either wired, employing fiber optic cables, or wireless, using cellular, satellite, Wi-Fi communication. The selection of the communication medium and frequency band is contingent on several factors including the geographical region, the specific application, and the volume of data to be transmitted. In particular, the 5G network, with superior attributes such as its high data capacity, low latency, and high reliability, is well positioned to manage the torrent of data generated by ITSs, facilitating swift delivery to the central system. The central system represents the hub where data from disparate APs are accumulated, processed, and converted into actionable decisions. Quick data processing is essential in ITSs to ensure timely actions based on the collected information. This requirement becomes of paramount importance as the trove of data produced by vehicles, infrastructure, and devices continues to grow exponentially. As a result, there is an escalating demand for efficient and high-capacity backhaul networks capable of processing this burgeoning data load effectively and expeditiously. In terms of network architecture, backhaul communications often adopt a hierarchical structure, with multiple layers of networks catering to different areas of coverage. At the highest level, the core network connects large geographical regions, while at lower levels, metropolitan area networks (MANs) and local area networks (LANs) provide connectivity in smaller areas. This tiered network design helps to manage network traffic effectively and reduce latency. Furthermore, quality of service (QoS) parameters are essential in managing the varied data traffic and ensuring that critical data are prioritized. Techniques such as packet scheduling and traffic shaping are employed to manage network resources efficiently.

As we move toward a future with more connected devices and vehicles, the role of backhaul communications in ensuring effective and efficient operation of ITSs becomes increasingly crucial. Technological advancements and better network management strategies promise to further enhance the capabilities of backhaul communications in the years to come.

One possible example that illustrates the utilization of interconnected networks to enhance EER is illustrated in Fig. 3. This figure illustrates a connected autonomous vehicle (CAV) concept where a platoon of multiple vehicles travels closely together and is controlled by a lead vehicle to maintain safety and efficiency. The lead vehicle is connected to the rest of the platoon vehicles through V2V links and receives information about the surrounding environment through V2I links. With appropriate processing of the received data, it is possible to control the positions and velocities of all of the platoon vehicles, which can ensure safety, efficiency, and enhanced EER. More specifically, it was observed

**Figure 3.** Wireless resource allocation for energy-efficient CAVs, cutting congestion, and $CO_2$ emissions.

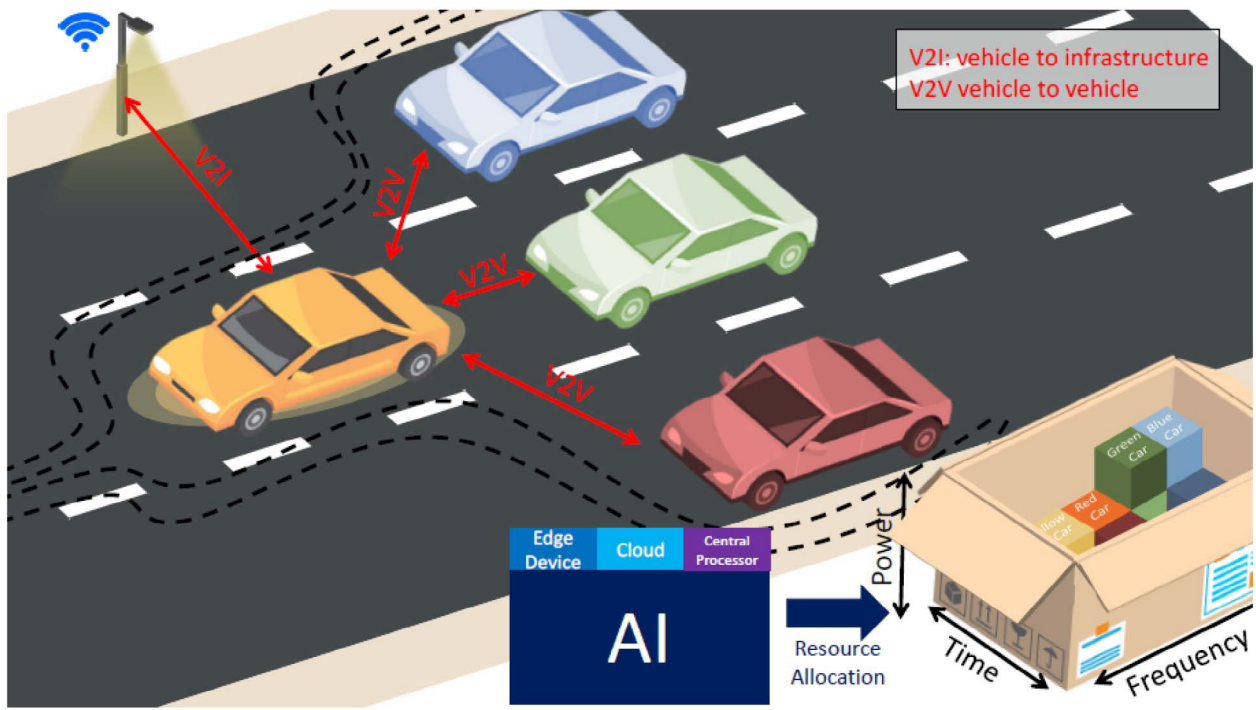

*Source: Authors.*

that driving in such a formation can exploit aerodynamic drafting (Liang, Mårtensson, and Johansson 2016). When vehicles travel closely together, the lead vehicle breaks the air resistance for the following vehicles, which then need less energy to maintain the same speed. This reduced need for energy can result in significant fuel savings and emission reduction. Achieving a significant EER enhancement requires an optimal optimization of interconnected network resources, where a decision-making algorithm takes the form of a convex/nonconvex optimization problem. The decision-making algorithm takes the received signals from the V2V/V2I communications and allocates wireless resources (i.e., frequency, time, and power) for each vehicle as an output. However, due to the complexity, stochastic nature, and high dynamics of the ITS context, convex/nonconvex optimization as decision makers might result in a suboptimal solution that can delay the communications

**Figure 4.** Platoon formations.

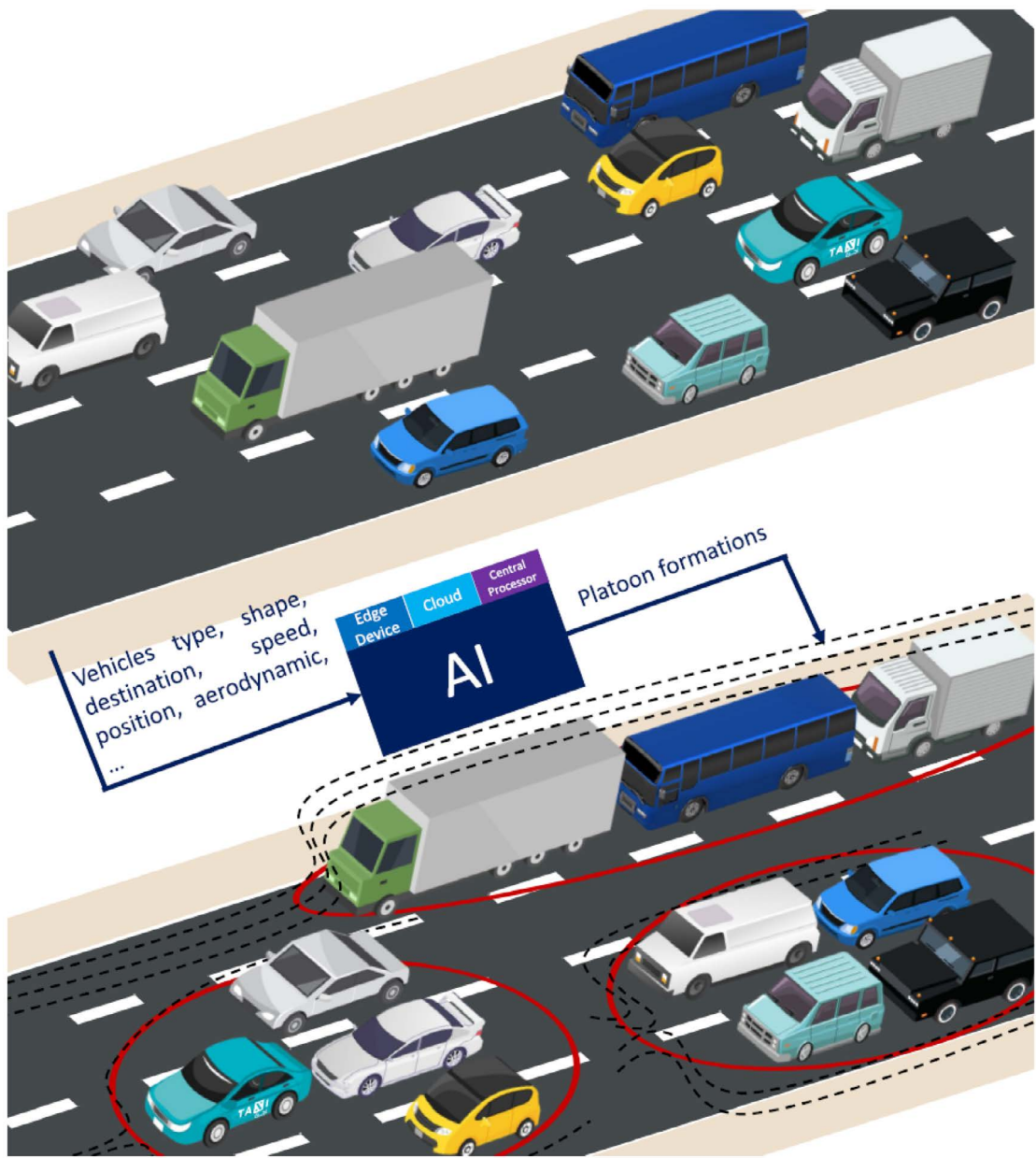

*Source: Authors.*

between the ITS elements, which can lead to an inappropriate control of passions and velocities of the platoon vehicles, which can risk safety and EER efficiency. Therefore, new AI-based solutions are emerging to tackle this problem. For instance, reinforcement learning (RL)-based AI has the capacity to control multiple agents (i.e., vehicles) with appropriate robustness against system dynamics and stochasticity.

The role of AI in this example goes beyond the resource allocation of communication links to further include the optimal platoon formation of multiple vehicles on the road. As shown in Fig. 4, a typical scenario includes a road with multiple vehicles of various shapes, sizes, destinations, speeds, and positions. In this context, AI receives this information as inputs in addition to other environmental/infrastructure parameters (e.g., airspeed, air direction) and computes the optimal platoon formations between the available vehicles with their typologies in a way that ensures optimal fuel efficiency. Moreover, interoperability emerges as a critical requirement in ensuring the efficient functionality of ITSs. It exemplifies the principle of interconnectivity, enabling diverse systems and devices – including traffic control systems, vehicle systems, and public transport systems – to communicate and exchange information seamlessly. This intercommunication concerns not only linkage but also ensuring that these different entities can understand and utilize shared information effectively, thereby enhancing overall network coordination and management. The result is an optimized traffic flow, heightened safety, and improved system efficiency. Beyond these immediate benefits, interoperability ensures that the ITS is scalable and adaptable, accommodating the smooth integration of new systems and technologies as transportation networks evolve. Interoperability is a fundamental prerequisite that guarantees an integrated, efficient, and future-ready urban mobility system, unlocking the full potential of ITSs.

# Promoting Enhanced EER in ITSs through AI

As ITSs continue to evolve, the inherent complexities and uncertainties within these systems also increase. As a result, various reasons can limit the effectiveness of the current ITS architectures. For instance, (i) traditional algorithms may not be able to process and analyze the vast amount of data generated in ITSs in a timely manner due to their limited computational efficiency. (ii) Traditional algorithmic approaches often struggle with uncertainty and incomplete information, which are common issues in complex, dynamic systems such as ITSs. (iii) Conventional methods often follow predetermined rules and cannot adapt to new or unforeseen circumstances. (iv) Traditional statistical methods may not have the predictive power necessary for many ITS applications, such as predicting traffic congestion or public transport demand. (v) Conventional ITS methods often apply the same rules or treatments to all users, ignoring individual differences. (vi) Traditional algorithms often struggle to integrate and interpret multimodal data – information from different sources and in different formats. Finally, (vii) traditional methods may not effectively optimize resource allocation, leading to inefficiencies such as buses running with few passengers or traffic congestion at intersections.

As a result, AI, with its ability to learn from large amounts of data, handle complex systems, and adapt to new situations, is uniquely positioned to address these challenges. AI techniques such as ML can model complex behaviors and patterns without needing explicit programmatic rules, making them effective in dealing with the dynamic and stochastic nature of ITSs. Furthermore, these techniques can generalize from learned data to new, unseen situations, making them robust to the varied and changing conditions in ITSs. Moreover, as previously mentioned, RL is specifically well suited to handle the optimization problems that lie at the heart of improving energy efficiency and reducing emissions (Pandharipande et al. 2023). RL can learn from interaction with the environment to make sequences of decisions that maximize a reward function, such as minimizing energy use or emissions. Importantly, it can do so without a model of the environment, making it an effective tool in situations where building an accurate model is difficult or impractical. AI's prowess extends beyond predictive capabilities. It is also instrumental in automating decision-making processes, optimizing resource allocation, and enabling real-time adaptability, all of which are crucial for managing and operating next-generation ITSs. This section explores how AI is harnessed across various facets of ITSs, from traffic management and fleet management to AV operation, predictive maintenance, smart grids, eco-driving systems, and emission monitoring and prediction. Each of these areas showcases AI's instrumental role in enhancing energy efficiency, reducing emissions, and paving the way toward a sustainable future.

## A. AI in Traffic Management and Optimization

Traffic management and optimization encompass the strategies and technologies employed to enhance the flow of traffic on roadways, mitigate congestion, and boost overall transportation efficiency. In today's bustling urban environments, traffic congestion is a pervasive issue, causing significant time loss, escalated fuel consumption, and heightened levels of air pollution.

AI has brought a revolutionary shift in this arena, offering advanced, adaptive, and scalable solutions to these challenges. Historically, traffic management was largely dependent on predefined rules and schedules.

However, these approaches often falter when confronted with the dynamic nature of traffic, which is influenced by a multitude of factors, including peak hour congestion, road accidents, weather conditions, and special events. The capabilities of AI, which has the ability to ingest and process vast volumes of data from diverse sources such as road sensors, traffic cameras, GPS tracking from vehicles, and social media feeds, become increasingly significant in this context. Harnessing ML algorithms, AI can learn complex traffic patterns, predict congestion levels, and adapt traffic control measures in real time. One of the most common AI techniques employed in this context is RL. In the context of traffic management, the traffic system constitutes the environment, the agent may represent a traffic signal controller, and the reward could be defined as a combination of elements such as minimizing the average delay, the number of stops, and queue lengths.

AI's contributions to traffic management are multifaceted:

- Traffic signal control: AI can be employed to dynamically optimize traffic signal timings. In this regard, a real-world example is Surtrac (Smith et al. 2013), an adaptive traffic control system deployed in Pittsburgh, U.S. The system uses AI to adaptively time traffic signals based on actual incoming traffic, as observed through cameras mounted on signal masts. Surtrac has successfully reduced travel times by 25%, wait times by 40%, and emissions by 20%.

- Traffic prediction: AI algorithms can predict traffic congestion by analyzing historical and real-time data, thereby enabling proactive traffic management. A case in point is Google Maps, which leverages ML to predict traffic and suggest the quickest routes to users (Derrow-Pinion et al. 2021).

- Incident detection and management: AI can rapidly detect and respond to incidents such as accidents or roadwork by analyzing data from cameras and sensors. Doing so facilitates real-time traffic rerouting and quick response from emergency services (Olugbade et al. 2022).

- Public transportation management: AI can optimize public transportation schedules based on predicted passenger demand. This optimization enhances the efficiency of public transport and promotes its use over private transport, thereby reducing overall emissions (Abduljabbar et al. 2019).

## B. AI in Fleet Management and Route Planning

Fleet management and route planning encapsulate the suite of strategies and technologies designed to effectively manage a fleet of vehicles and optimize their routes for efficiency. The fleet in question may range from delivery trucks and taxi services to public buses and AV groups. The effective management of such fleets is crucial for reducing operating costs, enhancing service reliability, and minimizing environmental impact (Li et al. 2015). Therefore, the integration of AI in this domain presents significant opportunities for efficiency gains and emission reduction. In the past, fleet management relied on relatively static route planning, with little room for real-time adaptation based on traffic conditions, weather, the status of vehicles, or other dynamic factors. This approach, however, can lead to inefficiencies, such as increased travel times, fuel consumption, and carbon emissions. AI as an alternative can utilize historical data and real-time inputs from a variety of sources, such as GPS trackers, traffic reports, and weather forecasts, to generate highly optimized and adaptive route plans. ML algorithms, including deep learning (DL) and RL models, are commonly employed to predict traffic patterns, calculate the optimal order of stops, and dynamically adjust routes based on changing conditions. For instance, consider a delivery company that uses AI to optimize the routes of its delivery trucks. By incorporating variables such as delivery locations, traffic and weather conditions, vehicle capabilities, and even driver work hours, the AI system can generate optimal routes that minimize the total travel distance and time, thereby reducing fuel consumption and emissions. Moreover, the AI system can adjust these routes in real time as conditions change, for example, if a road is closed due to construction or if an additional delivery is added on short notice.

A real-world example of this integration of AI in action is the routing algorithms used by ride sharing companies such as Uber and Lyft. These algorithms continuously analyze data on traffic, demand, and driver locations to efficiently match drivers with riders and generate optimal routes in real time. Moreover, AI can assist in other aspects of fleet management beyond route planning. For example, predictive maintenance can be significantly enhanced through ML models, allowing for potential issues to be identified and addressed before they result in costly downtime. AI can also assist with efficient fleet scheduling, ensuring that the number of active vehicles aligns with demand to avoid unnecessary fuel consumption and emissions (Liang et al. 2021; Shi et al. 2019; Theissler et al. 2021).

## C. AI in AVs

AVs, also known as self-driving cars, represent one of the most significant advancements in ITSs. These vehicles are designed to navigate and operate in their environments without human intervention. The realization of fully autonomous vehicles has the potential to dramatically transform our transportation systems, improving road safety, efficiency, and sustainability. The development and operation of AVs heavily depend on AI technologies, and their role in energy efficiency and emission reduction is multifaceted. At the core of AVs lies the AI system, which essentially acts as the "brain" of such a vehicle. It processes data from various sensors (such as LiDARs, RADAR, and cameras), creates a perception of the environment, makes decisions, and controls the vehicle's actions. These tasks demand sophisticated AI techniques, with DL being prominent due to its superior capability in handling complex tasks such as object detection, recognition, and tracking, as well as scene understanding.

Beyond basic navigation and control, AI in AVs can contribute to energy efficiency and emission reduction in several ways. (i) The first way is optimal path planning. Using AI, AVs can find the most efficient route to a destination, taking into account traffic conditions, the road type, and other factors. This optimized path planning can minimize the travel time and fuel consumption. (ii) The second is eco driving. AI enables AVs to adopt eco-driving strategies, such as smooth acceleration and deceleration, maintaining optimal speeds, and minimizing the idling time (Lakshmanan, Sciarretta, and Mourlan 2021). These strategies can significantly reduce fuel consumption and emissions. (iii) The third is platooning. AVs can form platoons where multiple vehicles travel closely together at high speed, reducing air resistance and, thus, energy consumption (Lakshmanan, Sciarretta, and Mourlan 2021). Platooning is made possible by V2V communication and AI algorithms that coordinate the vehicles. (iv) The fourth is predictive maintenance. ML models can predict maintenance needs based on vehicle usage and sensor data, allowing for issues to be fixed before they lead to higher fuel consumption or emissions. (v) The fifth and final way is improved traffic flow. By reducing human errors, which are a major cause of traffic congestion, AVs can improve overall traffic flow and efficiency, further reducing emissions.

## D. AI in Predictive Maintenance

Predictive maintenance stands as a key application of AI in the realm of ITSs. It revolves around the use of AI technologies to predict equipment failures before they happen, based on various indicators such as vibration, temperature, and pressure. Through such prediction, maintenance can be scheduled just in time to prevent equipment failures, enhancing overall operational efficiency, reducing downtime, and decreasing maintenance costs (Çınar et al. 2020). Conventionally, maintenance has been carried out either reactively, when a failure occurs, or preventively, based on a predetermined schedule. Both of these strategies have significant drawbacks. Reactive maintenance can lead to unexpected downtime and potentially high repair costs, while preventive maintenance can result in unnecessary work if the equipment does not need servicing. Predictive maintenance, fueled by AI, presents a better approach, aligning maintenance activities with the actual status and performance of equipment. AI's ability to learn from historical data and recognize patterns that may indicate an impending failure is the critical enabler of predictive maintenance. ML, particularly techniques such as DL and anomaly detection, is extensively used for this purpose.

In the context of ITSs, predictive maintenance can be applied to various components, such as vehicles, infrastructure (e.g., bridges, roads, traffic signals), and equipment (e.g., sensors, cameras). For example, in a vehicle, an AI-powered system could monitor data from various sensors (engine temperature, oil pressure, vibration levels, and others) to predict potential mechanical issues. This monitoring allows for proactive repairs, preventing an issue from escalating into a major failure that could lead to more significant energy consumption, emissions, or even safety risks.

## E. AI in Eco-Driving Systems

Eco-driving systems represent an innovative application of AI within the realm of ITSs. Essentially, these systems utilize AI to guide drivers or control vehicles in a manner that optimizes fuel efficiency and that reduces emissions without compromising on safety or travel time. Conventional driving behavior is often characterized by inefficient practices such as abrupt braking, rapid acceleration, and excessive idling. These behaviors not only increase fuel consumption and emissions but also contribute to traffic congestion and wear and tear on vehicles. Eco-driving systems aim to counteract these inefficiencies by leveraging AI to provide real-time feedback to drivers or to control AVs for the most eco-efficient driving behavior (Delnevo et al. 2019).

The role of AI in eco-driving systems is twofold: (i) predictive modeling and (ii) control optimization. AI is used to predict the future state of the vehicle and its environment based on sensor data, while optimization algorithms determine the best control actions to minimize energy consumption. In predictive modeling, AI models are used to predict the vehicle's future state and its environment. This prediction includes forecasting upcoming road conditions (such as traffic lights, congestion, and the road gradient), predicting the vehicle's speed and acceleration, and estimating fuel consumption and emissions based on driving behavior. In control optimization, based on the predictions, AI algorithms determine the optimal driving behavior to minimize fuel consumption and emissions. This determination might include the optimal acceleration and deceleration, speed, gear shifting (for manual vehicles), and idling behavior. For AVs, AI can directly control such vehicles based on these optimal actions. For human drivers, the system can provide real-time feedback and suggestions to guide eco-efficient driving. Various ML techniques, including regression, neural networks, and RL, are used to create predictive models and optimization algorithms in eco-driving systems. For instance, RL can be used to create a policy that maximizes fuel efficiency under various driving conditions, while neural networks can be used to predict future traffic conditions based on real-time and historical data (Guo et al. 2019; Lee et al. 2020).

## F. AI in Emission Monitoring and Prediction

The monitoring and prediction of emissions form a vital aspect of any strategy aimed at reducing the environmental impact of transportation systems. AI plays a key role in this context by enabling the accurate quantification of current emissions and the prediction of future emission trends. These capabilities are crucial for policy formulation, system optimization, and environmental protection. AI models can be trained on a vast array of data, including traffic flow characteristics, vehicle types, weather conditions, driving behaviors, and sensor readings, to create comprehensive emission models. These models can then provide detailed emission inventories, identify high-emission zones and times, and forecast future emission scenarios under different conditions and policies. The techniques used in these models can range from traditional ML methods such as linear regression and decision trees to more complex DL approaches, depending on the complexity of the task and the available data. In terms of emission prediction, AI can be instrumental in forecasting the impacts of different transportation strategies on emission levels. This forecasting can include the implementation of new technologies, changes in traffic management practices, shifts in public transportation usage, or the introduction of new policies. Such predictions can inform decision making and help in designing strategies that maximize emission reduction (Chavhan et al. 2022).

# Advancing Intelligent Transportation: The Saudi Arabian Experience and Vision

Amid burgeoning urbanization and increasing demands on its transportation infrastructure, Saudi Arabia faces unique ITS challenges with regard to conserving energy and reducing emissions. Population growth and urban sprawl have heightened the necessity for sophisticated and efficient traffic management systems. Moreover, as a traditionally fossil fuel-dependent nation, Saudi Arabia recognizes the urgent need for sustainable solutions to lessen environmental impacts, stimulate economic growth, and pave the way for a sustainable future. In response, the Saudi government has spearheaded several groundbreaking initiatives aimed at revolutionizing its transportation landscape. A cornerstone of this transformation is the National Industrial Development and Logistics Program (NIDLP), designed to transform Saudi Arabia into a leading logistics hub, primarily by integrating diverse transport modes with cutting-edge technologies. Smart traffic management projects, which are a crucial component of this transformation, employ advanced surveillance systems, intelligent traffic signals, and real-time traffic management centers. Saudi Arabia's ambitions are further exemplified by the audacious NEOM project (Alam et al. 2021). Positioned to be a model for the cities of the future, NEOM is envisioned to be a car-free zone. Instead of traditional vehicles, the city plans to utilize an array of AVs, drones, and other advanced mobility solutions, supporting a framework where every destination within the city is easily reachable without the need for private car ownership. This paradigm shift underscores Saudi Arabia's commitment to fostering sustainable, efficient, and intelligent urban mobility. Moreover, significant strides have been made toward minimizing energy consumption and emissions in the transport sector. Under Saudi Vision 2030, the nation is promoting the adoption of electric and hybrid vehicles as part of its sustainable transport strategy. Additionally, substantial investments in public transportation infrastructure, such as the Riyadh Metro and Jeddah Public Transport Program, aim to reduce the dependency on private vehicles, thereby contributing to energy conservation and reduced emissions.

These initiatives carry far-reaching implications not only for environmental sustainability but also for economic growth and tourism. Enhanced transportation efficiency boosts Saudi Arabia's logistical competitiveness, supporting various sectors and aiding in economic diversification. Moreover, improved public transportation infrastructure and services contribute to Saudi Arabia's allure as a tourist destination by providing convenient access to cultural and heritage sites.

Looking forward, Saudi Arabia is poised to emerge as a frontrunner in the adoption of ITSs and sustainable transportation practices. Its commitment to digital transformation, as evidenced by initiatives such as the Digital Transformation Unit and Saudi Data and AI Authority, suggests the potential for significant AI integration in the country's ITS. Such integration is expected to revolutionize traffic management, safety, and user experience. Through strategic planning, effective implementation, and continued innovation, Saudi Arabia is charting a course toward becoming an exemplary figure in the realm of intelligent, sustainable, and inclusive transportation.

# Conclusions

In this paper, we presented a comprehensive examination of ITSs, specifically their potential to enhance EER through the deployment of AI. The discussion began with an exploration of the integral role of sensors in creating environmentally friendly ITS services. Whether embedded in vehicles, forming part of the transportation infrastructure, or integrated into devices, these sensors are key enablers in gathering crucial data to facilitate intelligent decision making and enhance energy efficiency. The subsequent section delved into the interconnected networks of ITSs, highlighting how V2V, V2I, V2C, and backhaul networks serve as critical communication links. These networks underpin the functioning of ITSs, playing a pivotal role in energy conservation and emission reduction by improving traffic management and efficiency. Building upon the foundations of sensor technologies and interconnected networks, we investigated the transformative potential of AI in further enhancing EER in ITSs. This analysis spanned multiple facets of ITSs, such as traffic management and optimization, fleet management and route planning, AVs, predictive maintenance, eco-driving systems, and emission monitoring and prediction. Each of these areas stands to gain significant improvements in energy efficiency and emission reduction through AI integration.

In particular, the case of Saudi Arabia offers intriguing insights into the ambitious efforts underway to harness the power of ITSs and AI for sustainable transportation. The Saudi government's initiatives, such as the NIDLP and the vision of the futuristic city of NEOM, underscore its commitment to a transformative approach in addressing energy and environmental challenges.

These efforts embody the belief that an ITS empowered by AI can serve as a pivotal tool in achieving significant EER, driving economic growth, and enhancing the quality of life for Saudi citizens. The journey toward a sustainable transportation future is multifaceted, requiring the integration of advanced technologies, strategic planning, and collaborative effort. As we move forward, it is clear that the convergence of ITSs, AI, and sustainability principles offers promising avenues to address our energy and environmental challenges. The ongoing endeavors in Saudi Arabia present a beacon of what can be achieved and are a testament to the transformative power of the ITS-AI-EER triad. This report concludes with the anticipation that Saudi Arabia's vision and actions will continue to inspire and guide other nations in their journey toward a sustainable and intelligent transportation future.

# About the Project

The project aims to create an evidence-based, comprehensive understanding of the different projects, policies and regulations around international connectivity, regional transport and urban mobility in Saudi Arabia, their contribution to energy demand and lowering GHGs and emissions, as well as setting the Kingdom on a net-zero pathway.

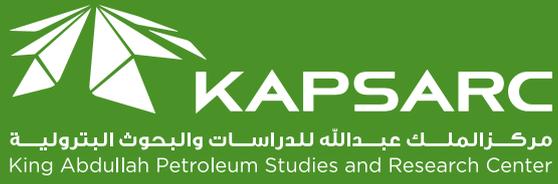

www.kapsarc.org